\documentstyle[epsf,multicol,aps]{revtex}

\begin{document}
\draft
\title {Laser Field Induced Birefringence and Enhancement of Magneto-optical 
Rotation\footnote{An article in honor of Prof. M. O. Scully - who led us into 
many areas of quantum optics and laser physics.}}

\author {
Anil K. Patnaik$^1$ and G. S. Agarwal$^{1,2}$}
\address {$^1$Physical Research Laboratory, Navrangpura, Ahmedabad-380 009, 
India\\
$^2$Max-Plank-Institut f\"{u}r Quantenoptik, 85748 Garching, Germany}
\date{June 28, 1999}

\maketitle 
\begin{abstract} 
An initially isotropic medium, when subjected to either a
magnetic field or a coherent field, can induce anisotropy in the medium and can 
cause the polarization of a probe field to rotate. Therefore the rotation of
probe polarization, due to magnetic field alone, can be controlled 
efficiently with the use of a coherent control field. We demonstrate this 
enhancement of the magneto-optical rotation (MOR) of a linearly polarized 
light, by doing detailed calculations on a system with relevant transitions
$j=0\leftrightarrow j=1\leftrightarrow j=0$.
\end{abstract}

\section{Introduction}

An isotropic medium having $m$-degenerate sub-levels when subjected to a
magnetic field exhibits birefringence in its response to a polarized optical
field. This is due to the fact that Zeeman splitting of magnetic sub-levels
causes asymmetry in refractive indices for left and right circular polarization
components of the optical field. The result is magneto-optical rotation (MOR);
i.e., the plane of polarization of the light emerging out of the medium is 
rotated with respect to that of the incident. If $\chi_+(\chi_-)$ represents 
the susceptibility of the birefringent medium corresponding to right 
(left) circular polarization component of the probe, the rotation angle is
given by
\begin{equation}
\theta = \pi k_pl(\chi_- - \chi_+),
\end{equation}
where $\vec{k}_p$ corresponds to propagation vector of the probe and $l$ is 
length of the 
medium along the direction of propagation. The susceptibilities $\chi_\pm$ 
are functions of characteristics of atomic transitions and hence studies of 
$\theta$ yield information on atomic transitions.
Extensive literature on MOR exists. Experiments have been carried out
with both synchrotron and laser sources \cite{synch,laser,pol-book}. 
Effects of larger probe powers and resulting saturation have also been studied 
at length \cite{sat-gawlik,sat-gsa,sat-scully,sat-cohen}.

	The present study is motivated by the possibility that the 
susceptibilities $\chi_\pm$ can be manipulated by the application of a control
laser \cite{chi-gsa,chi-harris,chi-scully,chi-xiao}. The control laser 
modifies the atomic structure and thus by
tuning the strength and the frequency of the control laser one can obtain
suitable modifications in $\chi_\pm$. Wielandy and Gaeta \cite{gaeta} recently 
demonstrated control of polarization state of the probe field in an 
initially isotropic medium (see also \cite{biref,rot}). In this paper we study 
in detail the possibility of the enhancement of MOR produced by the control 
laser. 

	The organization of the paper is the following: in Sec {\rm II}, we
describe the response of an isotropic medium to a linearly polarized weak 
probe. In Sec.{\rm III}, we calculate $\chi_\pm$ of the medium in presence of
a control field using the density matrix formalism. We show analytically how
a circularly polarized control field causes anisotropy in the medium. Further
in in Sec {\rm IV}, we present the numerical results - that demonstrate the
control laser induced birefringence and we show the resulting large 
enhancement in MOR in the regions identified by the analytical results of
Sec.{\rm III}. We also show new probe-frequency domains where significant MOR
can occur. 

\section{Response of an anisotropic medium to a linearly polarized light}
	
	Let us consider an incident probe field $\vec{E}_{in}$ with frequency 
$\omega_p$ propagating inside an active medium along the 
quantization axis $z$ is 
\begin{equation}
\vec{E}_{in} (z,t) = {\vec {\cal E}}_p (z) e^{-i\omega_p t + ik_pz} + c.c.~ .
\end{equation}
We resolve the incident field amplitude ${\vec {\cal E}}_p$ into two 
circularly polarized $\sigma_+$ and
$\sigma_-$ components 
\begin{equation}
{\vec{\cal E}}_p = \hat{\epsilon}_+ {\cal E}_{p+} + \hat{\epsilon}_{p-} 
{\cal E}_{p-},
\end{equation}
where unit polarization vectors $\hat{\epsilon}_\pm$ corresponding to 
$\sigma_\pm$ polarization are defined in terms of unit
vectors $\hat{x}$ and $\hat{y}$ as
\begin{equation}
\hat{\epsilon}_\pm = \frac{\hat{x} \pm i \hat{y}}{\sqrt 2}~.
\end{equation}
The polarization induced in the atomic medium due to a linearly polarized
probe field can be expressed as
\begin{equation}
\vec{{\cal P}} = \left( \hat{\epsilon}_+ \chi_+ {\cal E}_{p+} + 
\hat{\epsilon}_- \chi_- {\cal E}_{p-}\right) e^{-i\omega_p t + ik_pz} + c.c.~ ,
\end{equation}
where $\chi_\pm$ gives response of the medium corresponding to 
$\sigma_\pm$ component of the electric field and, in general, is function 
of the fields ${\cal E}_{p\pm}$. In a previous work Agarwal {\it et al} 
\cite{sat-gsa} studied
the MOR under condition of saturation. However here we consider propagation 
of a weak probe and thus $\chi_\pm$ are independent of ${\cal E}_{p\pm}$. In 
the steady state the solution of the wave equation for ${\cal E}_{p\pm}$ gives 
the output field amplitude
\begin{equation}
\vec{\cal E}_{out}(z=l) = \hat{\epsilon}_+ {\cal E}_{p+}(0) e^{2\pi ik_pl\chi_+}
+ \hat{\epsilon}_- {\cal E}_{p-}(0) e^{2\pi ik_pl\chi_-}.
\end{equation}
For an $x$-polarized incident field say $\vec{\cal E}_{in} = \hat{x} {\cal E}_0$,
\begin{equation}
{\cal E}_{p+}(0) = {\cal E}_{p-}(0) =\frac{{\cal E}_0}{\sqrt{2}}, 
\end{equation}
the output field in Eq.(6) takes the form
\begin{equation}
\vec{\cal E}_{out} = \frac{{\cal E}_0}{\sqrt{2}}\left(
\hat{\epsilon}_+ e^{2\pi ik_pl\chi_+} + \hat{\epsilon}_- e^{2\pi ik_pl\chi_-}
\right).
\end{equation}
Thus the output field $\vec{\cal E}_{out}$ consists of both
$x$ and $y$-polarization components. The direction of light polarization 
rotates with respect to the polarization of the incident light. Experimentally 
the rotation is measured by measuring the intensity of transmission through a 
crossed polarizer at the output. In the above case the intensity of 
transmission through a $y$-polarized analyzer is given by 
\begin{equation}
T_y = \frac{|(\vec{\cal E}_{out})_y|^2}{|\vec{\cal E}_{in}|^2} =
\frac{1}{4}\left| e^{2\pi ik_pl\chi_+} - e^{2\pi ik_pl\chi_-} \right|^2, 
\end{equation}
where the output intensity is normalized with respect to the input.
It may be noted that for a resonant or near-resonant probe field, $\chi_\pm$ 
will be complex and therefore due to large absorption by the medium, there
will be large attenuation of output signal. However assuming that $\chi_\pm$ 
is real, 
i.e. absorption is small, we get from Eq.(8)
\begin{equation}
\frac{{\cal E}_y}{{\cal E}_x} = \tan \theta = \tan \pi kl Re(\chi_- - \chi_+).
\end{equation}

	For a commonly studied transition $j=0\leftrightarrow j=1$, the above
susceptibilities are given by 
\begin{equation}
\chi_\pm = \left( \frac{\alpha}{4\pi k} \right)\frac{\gamma}{(\delta\pm\Omega 
-i\gamma)},
\end{equation}
where $2\hbar\Omega$ is the Zeeman splitting between the levels 
$|j=1,m=1\rangle$ and $|j=1,m=-1\rangle$; $\delta$ is detuning between the 
the probe frequency and the
frequency $\omega_{10}$ of the transition $j=1$ to the ground state $j=0$ 
(with zero magnetic field) 
\begin{equation}
\delta = \omega_{10} - \omega_p.
\end{equation}
In Eq.(11), $2\gamma$ is the decay rate of say $|j=1,m=\pm 1\rangle$ to the 
level $|j=0,m=0\rangle$. The quantity $\alpha l$ gives the resonant absorption
\begin{equation}
\alpha = \frac{4\pi k |d|^2 n}{\hbar\gamma}, 
\end{equation}
where $n$ is the density of atoms and $d$ is dipole matrix element for the 
transition. It is to be noticed that $\chi_+ = \chi_-$ for zero magnetic field.
Clearly to produce large MOR we have to make $\chi_\pm$ differ from each other
to maximum possible extent. In next section we calculate the effect of a 
control laser on $\chi_\pm$.

\section{Calculation of $\chi_\pm$ in presence of a control laser}

We consider a model system [see Fig.1] involving say cascade of transitions
$|j=0,m=0\rangle$ (level $|g\rangle$) $\leftrightarrow ~|j=1,m=\pm 1\rangle$
(level $|1\rangle$ and $|2\rangle$) $\leftrightarrow ~|j=0,m=0\rangle$ (level
$|e\rangle$). This for example will be relevant for expressing $^{40}Ca$. The 
probe $\vec{E}_p$ will act between the levels $|g\rangle$ and $|1\rangle, 
|2\rangle$. We assume in addition the interaction of a control laser 
$\vec{E}_c$ to be
nearly resonant with the transition $|e\rangle \leftrightarrow |1\rangle,
|2\rangle$. For simplicity we drop the transition $m=0\leftrightarrow m=0$. We 
thus assume the loss to $m=0$ state by spontaneous emission could be pumped
back by an incoherent pump. Let $\omega_{\alpha\beta}$ be the transition 
frequency between the levels $|\alpha\rangle$ and $|\beta\rangle$. The total 
Hamiltonian of the atom interacting with the control and probe fields is
\begin{eqnarray}
{\cal H} &=&  \hbar \omega_{eg} |e\rangle\langle e| +
\hbar \omega_{1g} |1\rangle\langle 1| + \hbar \omega_{2g} |2\rangle\langle 2|
\nonumber \\
& & - \vec{d}\cdot (\vec{E}_c + \vec{E}_p),
\end{eqnarray} 
where both $\vec{E}_c$ and $\vec{E}_p$ are given by Eqs.(2) and (3). We make 
rotating wave approximation and thus we approximate the interaction part of 
the Hamiltonian by 
\begin{eqnarray}
-\vec{d}\cdot (\vec{E}_c +\vec{E}_p) \approx 
&-&G_1 |e\rangle\langle 1| e^{-i\omega_c t+ik_c z} 
-G_2 |e\rangle\langle 2| e^{-i\omega_c t+ik_c z}
\\ \nonumber
&-&g_1 |1\rangle\langle g| e^{-i\omega_p t+ik_p z} 
-g_2 |2\rangle\langle g| e^{-i\omega_p t+ik_p z} + H.c.,
\end{eqnarray}
where $2G$s and $2g$s represent Rabi frequencies of the control and the
probe laser -
\begin{equation}
G_i = \frac{\vec{D}_{ei}\cdot \vec{\cal E}_c}{\hbar},~
g_i = \frac{\vec{d}_{ig}\cdot \vec{\cal E}_p}{\hbar}.
\end{equation}
In order to proceed further we will make following transformations on the
off-diagonal elements of the density matrix - this transformation removes
all explicit dependence on the optical temporal and spatial frequencies
\begin{eqnarray}
\rho_{ei} = \tilde{\rho}_{ei} e^{-i\omega_c t+ik_c z},
\rho_{ig} = \tilde{\rho}_{ig} e^{-i\omega_p t +ik_p z},
\\ \nonumber
\rho_{eg} = \tilde{\rho}_{eg} e^{-i(\omega_c +\omega_p)t +i(k_c +k_p)z}.
\end{eqnarray}
In the following we drop tildes from the density matrix equation. However it 
should be borne in mind that the full density matrix in Schr\"{o}dinger picture 
is to be obtained by using Eq.(17). On introducing the various rates of 
spontaneous emission, we can write the equations for density matrix elements
as 
\begin{eqnarray}
\dot{\rho}_{ee} &=& -2(\Gamma_1 + \Gamma_2) \rho_{ee} + iG_1\rho_{1e}
-iG_1^*\rho_{e1} +iG_2\rho_{2e} -iG_2^*\rho_{e2} ,
\nonumber\\ 
\dot{\rho}_{e1} &=& -(\Gamma_1 + \Gamma_2 +\gamma_1 +i(\Delta -\Omega))
\rho_{e1} +iG_1 (\rho_{11} -\rho_{ee}) +iG_2\rho_{21} -ig_1^*\rho_{eg},
\nonumber\\ 
\dot{\rho}_{e2} &=& -(\Gamma_1 + \Gamma_2 +\gamma_2 +i(\Delta +\Omega))
\rho_{e2} +iG_2 (\rho_{22} -\rho_{ee}) +iG_1\rho_{12} -ig_2^*\rho_{eg},
\nonumber\\ 
\dot{\rho}_{eg} &=& -(\Gamma_1 + \Gamma_2 +i(\Delta +\delta)) \rho_{eg} +i G_1\rho_{1g}
+ i G_2\rho_{2g} -i g_1\rho_{e1} -i g_2\rho_{e2},
\nonumber\\ 
\dot{\rho}_{11} &=& 2\Gamma_1\rho_{ee} -2\gamma_1\rho_{11} +iG_1^*\rho_{e1}
-iG_1\rho_{1e} +ig_1\rho_{g1} -ig_1^* \rho_{1g},
\\ 
\dot{\rho}_{12} &=& -(\gamma_1 +\gamma_2 +2i\Omega)\rho_{12} +iG_1^* \rho_{e2}
+ig_1\rho_{g2} -iG_2\rho_{1e} -ig_2^*\rho_{1g},
\nonumber\\ 
\dot{\rho}_{1g} &=& -(\gamma_1 +i(\delta+\Omega))\rho_{1g} +ig_1(\rho_{gg}
-\rho_{11}) + iG_1^*\rho_{eg} -ig_2\rho_{12},
\nonumber \\ 
\dot{\rho}_{22} &=& 2\Gamma_2\rho_{ee} -2\gamma_2\rho_{22} +iG_2^*\rho_{e2}
-iG_2\rho_{2e} +ig_2\rho_{g2} -ig_2^* \rho_{2g},
\nonumber\\ 
\dot{\rho}_{2g} &=& -(\gamma_2 +i(\delta-\Omega))\rho_{2g} +ig_2(\rho_{gg}
-\rho_{22}) + iG_2^*\rho_{eg} -ig_1\rho_{21},
\nonumber\\ 
\dot{\rho}_{gg} &=& 2\gamma_1\rho_{11} + 2\gamma_2\rho_{22} +ig_1^* \rho_{1g}
-ig_1\rho_{g1} +ig_2^*\rho_{2g} -ig_2\rho_{g2},
\nonumber
\end{eqnarray}
where the detunings are defined by 
\begin{eqnarray}
\Delta = (\omega_{e1} - \omega_c +\Omega) = (\omega_{e2} - \omega_c -\Omega),
~\omega_{12} = 2\Omega,
\\ \nonumber 
\delta = (\omega_{1g} - \omega_p -\Omega) = (\omega_{2g} - \omega_p +\Omega). 
\end{eqnarray}
In Eq.(18) $2\Gamma_i$ ($2\gamma_i$) represents rate of spontaneous emission
from $|e\rangle \rightarrow |i\rangle$ ($|i\rangle \rightarrow |g\rangle$).
However in further calculations we assume $\gamma_1 = \gamma_2 = \gamma$ for
simplicity. The dipole matrix elements in (16) are given in terms of vectors 
(4) as 
\begin{eqnarray}
\vec{D}_{e1} &=& -D\hat{\epsilon}_+,~ \vec{D}_{e2} = D\hat{\epsilon}_-,
\nonumber \\
\vec{d}_{1g} &=& -d\hat{\epsilon}_-,~ \vec{d}_{2g} = d\hat{\epsilon}_+.
\end{eqnarray}
Here $D$ ($d$) denotes the reduced dipole matrix element corresponding to
$|e\rangle \leftrightarrow |i\rangle$ ($|i\rangle \leftrightarrow |g\rangle$)
transitions. Clearly we also have 
\begin{eqnarray}
G_1 = -\frac{D{\cal E}_{c-}}{\hbar}&,& 
G_2 = \frac{D{\cal E}_{c+}}{\hbar}, 
\\ \nonumber
g_1 = -\frac{d{\cal E}_{p+}}{\hbar}&,& 
g_2 = \frac{d{\cal E}_{p-}}{\hbar}. 
\end{eqnarray}
This should be kept in view to determine the component of circular polarization
that connects various transitions.

	We next calculate the susceptibilities using solutions of Eq.(18). The
induced polarization at frequency $\omega_p$ will be obtained from off-diagonal
elements $\rho_{1g}, \rho_{2g}$:
\begin{eqnarray}
\vec{\cal P} &=& n Tr (\vec{d}\rho) = n (\vec{d}_{g1}\rho_{1g} +\vec{d}_{g2} 
\rho_{2g} + c.c.)
\\ \nonumber
&=& n(\vec{d}_{g1}\rho_{1g} + \vec{d}_{g2} \rho_{2g}) e^{-i\omega_p t+ik_p z}
+ c.c.~.
\end{eqnarray}
The exponential factors in (22) come from the transformation in (17). As
indicated earlier, $\rho_{1g}$ and $\rho_{2g}$ will be computed to the lowest
order in the probe field. Defining 
\begin{equation}
\rho_{1g} = \left(\frac{g_1}{\gamma}\right) s^+,
\rho_{2g} = \left(\frac{g_2}{\gamma}\right) s^-,
\end{equation}
and using (20) and (21), we can write (22) in the form
\begin{equation}
\vec{\cal P} = (\chi_+{\cal E}_{p+}\hat{\epsilon}_+
+ \chi_-{\cal E}_{p-}\hat{\epsilon}_-)e^{-i\omega_p t+ik_p z} + c.c.,
\end{equation}
where 
\begin{equation}
\chi_\pm = \left( \frac{\alpha}{4\pi k}\right) s^\pm.
\end{equation}
We have been able to obtain analytical solutions for $s^\pm$ which we give 
below -
\begin{equation}
s^+ = \frac{i\gamma \left[ |G_2|^2 +{\bf (}\gamma +i(\delta
-\Omega){\bf )(}\Gamma_1 +\Gamma_2 +i(\Delta +\delta){\bf )}\right]}
{|G_2|^2 {\bf (}\gamma +i(\delta +\Omega){\bf )}
+{\bf (}\gamma+i(\delta -\Omega){\bf )}
\left[ |G_1|^2 +{\bf (}\gamma +i(\delta +\Omega){\bf )}
{\bf (}\Gamma_1 +\Gamma_2 +i(\Delta +\delta){\bf )}
\right]},
\end{equation}
\begin{equation}
s^- = \frac{i\gamma \left[ |G_1|^2 +{\bf (}\gamma +i(\delta
+\Omega){\bf )(}\Gamma_1 +\Gamma_2 +i(\Delta +\delta){\bf )}\right]}
{|G_1|^2 {\bf (}\gamma +i(\delta -\Omega){\bf )}
+ {\bf (}\gamma +i(\delta +\Omega){\bf )}
\left[ |G_2|^2 +{\bf (}\gamma +i(\delta -\Omega){\bf )}
{\bf (}\Gamma_1 +\Gamma_2 +i(\Delta +\delta){\bf )}
\right]}.
\end{equation}
These reduce to well known results in the absence of the control field 
$G_1 = G_2 = 0$. On substituting (26) and (27) in (9) we find the y-component
of the transmitted field 
\begin{equation}
T_y = \frac{1}{4}\left| \exp\left( i\frac{\alpha l}{2}s^+\right)
-\exp\left( i\frac{\alpha l}{2}s^-\right)\right|^2.
\end{equation}
As mentioned earlier, we choose the parameters in such a way that there is
maximum asymmetry between $\chi_+$ and $\chi_-$. An important case occurs
when, say, $G_2 = 0$ i.e. the control field is $\sigma_-$ polarized 
(${\cal E}_{c+} =0$, ${\cal E}_{c-} \ne 0$). Clearly $\chi_-$ 
or $s^-$ has the value in absence of the control field
\begin{equation}
s^- = \frac{i\gamma}{{\bf (}\gamma +i(\delta -\Omega){\bf )}};
\end{equation}
where as $\chi_+$ is strongly dependent on the strength and frequency of the
control field
\begin{equation}
s^+ = \frac{i\gamma{\bf (}\Gamma_1 +\Gamma_2 + i(\Delta + \delta){\bf )}}
{|G_1|^2 + {\bf (}\gamma +i(\delta+\Omega){\bf )(}\Gamma_1 +\Gamma_2 + i(\Delta + \delta){\bf )}}
.
\end{equation}
For large $|G_1|$, both real and imaginary part of $s^+$ will show 
Autler-Townes splitting. Note further that even when no magnetic field is 
applied $\Omega=0$,
\begin{equation}
s^- (\Omega =0) \ne s^+(\Omega =0), ~~{\rm for}~ |G_1| \ne 0.
\end{equation}
In this case we have control laser induced birefringence. A particularly 
attractive possibility is to consider the case when $|G_1|$ is large and
that we work in the regime of other parameters so that $|s^+| \ll |s^-|$.
Under such conditions we will have large MOR or large signal $T_y$. The
experiments of Wielandy and Gaeta \cite{gaeta} focus on the laser induced birefringence.

\section{Numerical results on laser induced birefringence and enhancement 
of magneto-optical rotation}

	In this section we present numerical results to demonstrate how the
MOR can be enhanced. Note that the parameter space is rather large and the
results will depend on the choice of $|G_1|$, $|G_2|$, magnetic field, control
laser detuning, and of course the probe laser detuning. We have carried out
the numerical results for a large range of parameters and present some 
interesting results below. In all the numerical results we scale all 
frequencies with respect to $\Gamma_i (=\gamma)$ and we take the absorption 
co-efficient $\alpha l$ as 30. 

	In Fig.2 we consider the rotation of plane of polarization in the 
absence of the magnetic field. For no control laser, the rotation vanishes. 
For non-zero $G_1$, the medium becomes anisotropic {\bf (}Eq.(31){\bf )}, we 
obtain substantial rotation of the plane of polarization. The Fig.2 also shows 
the real and imaginary part of $\chi_\pm$. In Fig.3, we show the results for
non-zero value of magnetic field. We find definite enhancement in the 
magneto-optical rotation. The reason for this enhancement can be traced back
to the {\it flipping of the sign of Re $\chi_+$ which is caused by the control 
field}. In addition we can produce large rotation for probe frequencies in 
the neighborhood of the frequencies of the Autler-Townes components. In Fig.3 
we also show how the non-resonant control field can produce further 
enhancement. Our calculations also suggest some interesting domain in which 
the probe should be tuned to obtain large MOR. Application of the control 
laser permits us to obtain large MOR in totally different frequency regime. 
In Fig.4 
we show the changes in the transmitted signal if $G_2\ne 0$. The absorption
and dispersion profile now exhibit a triplet structure. As noted earlier,
intensity of the transmitted y-component of the signal depends on the asymmetry
between $\chi_+$ and $\chi_-$. When asymmetry goes down, as for example,
in the third and fourth row (from ($a_2$) to ($e_2$)) of Fig.4, then $T_y$ 
can go down. In our model the control of $\chi_\pm$ is $not$ independent of 
each other as we connect to a common final level. Clearly a large enhancement 
will be possible if $\chi_\pm$ could be independently manipulated. 

	In conclusion we have shown how a control laser can produce 
birefringence as well as enhance the amount of magneto-optical rotation effect.
Besides the control field can produce new frequency regions which show 
very significant magneto-optical rotation.

	GSA thanks S. Stenholm for pointing out Ref.[14] and for discussions. 
GSA is also grateful to Marlan Scully for extensive discussions and insights 
on laser induced control of optical properties.

\vspace*{1cm}
\centerline{\bf FIGURES}
\vspace*{1cm}

\begin{figure}
\epsfxsize 3.3 in
\centerline{
\epsfbox{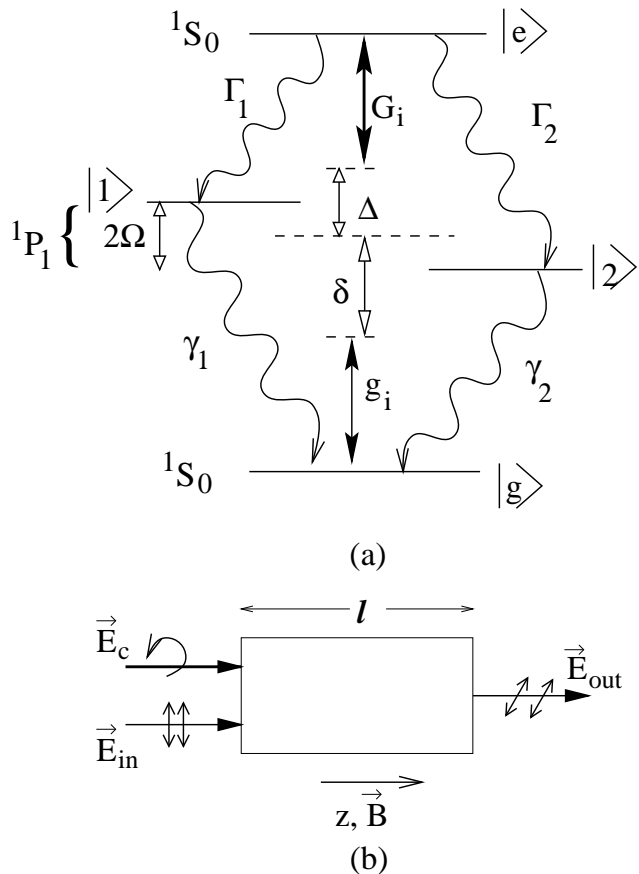}}
\caption{
(a) The four-level model scheme (say of $^{40}Ca$) having $m$-degenerate 
sub-levels as its intermediate states. The symbols in left hand side
denote the energy levels of $^{40}Ca$ atom. $2\Gamma_i$ and $2\gamma_i$ are
the spontaneous decays, $2g_i$ ($2G_i$) is Rabi frequency of the probe (control)
field due to coupling of the intermediate state $|i\rangle$ with $|g\rangle$
($|e\rangle$). The detuning of probe (control) field from the center of 
$|1\rangle$ and $|2\rangle$ are represented by $\delta$ ($\Delta$). 
$2\Omega$ is the Zeeman split between the intermediate states.
(b) A block diagram that shows the configuration under consideration. $\vec{B}$
defines the quantization axis $z$. The input probe $\vec{E}_{in}$ is
$x$-polarized and the control field is left circularly polarized. Both the 
fields propagate along $z$. After passing through the cell, output
is observed through a $y$-polarized analyzer.
}
\end{figure}

\vspace*{3cm}
\begin{figure}
\epsfxsize 5.3 in
\centerline{
\epsfbox{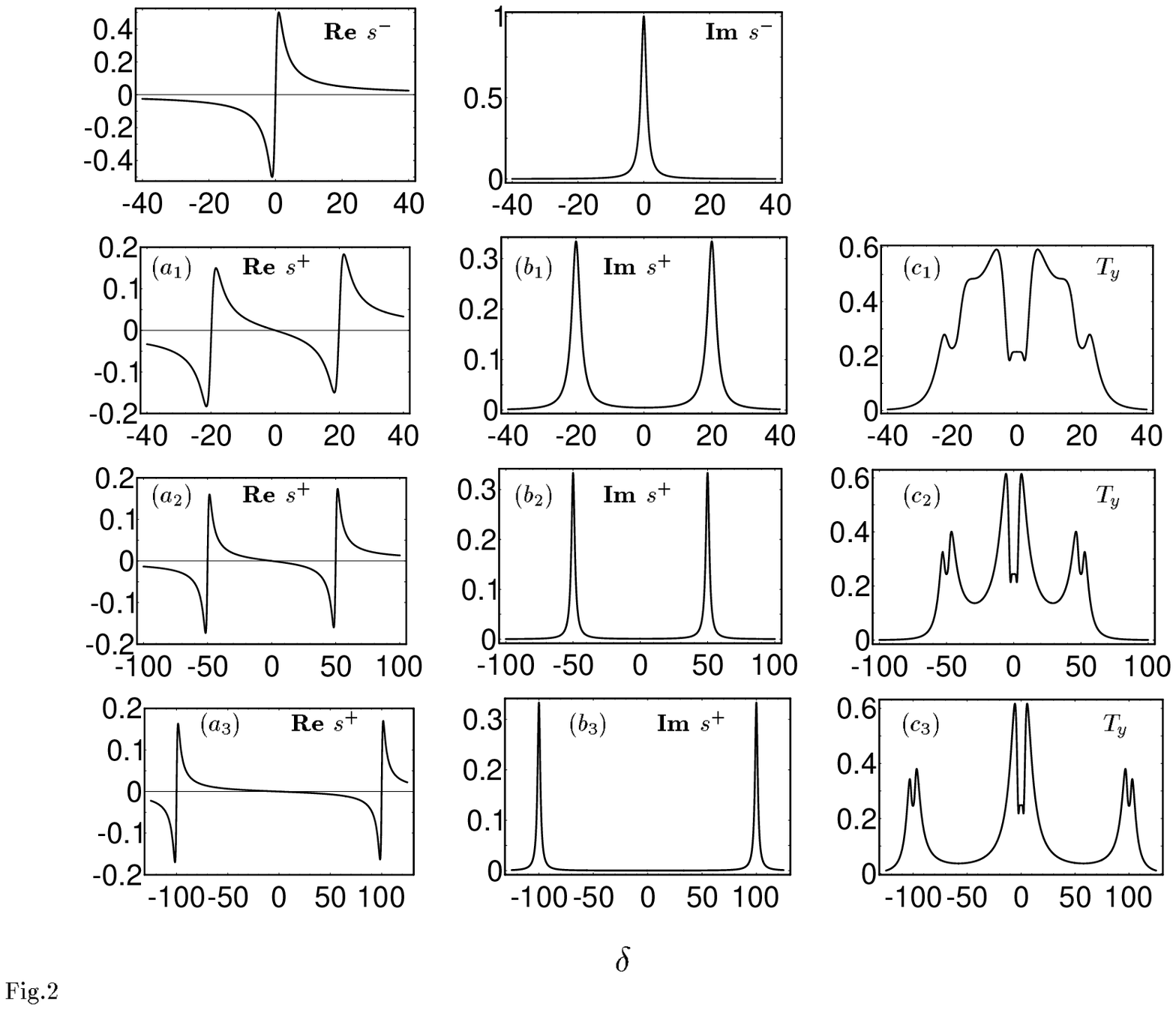}}
\vspace*{.5cm}
\caption{
Plots to show the control field induced birefringence in absence of
magnetic field ($\Omega = 0$). Control field is left circularly polarized 
and is in resonance with $|i\rangle \rightarrow |e\rangle$ transition 
($\Delta = 0$). In the above figure, first row refers to real and imaginary
part $s^-$ (Eq.(29)) which is independent of the $\sigma_-$ control field.
The frames
($a_j$), ($b_j$) and ($c_j$) represent Re $s^+$, Im $s^+$ and $T_y$ 
respectively. Three rows for $j=1,~2,~3$ correspond to control field strengths 
$G_1 = 20,~50,~100$ respectively. The x-axis represents the detuning of probe 
laser with respect to the position of the $j=1$ level with zero magnetic
field. 
}
\end{figure}

\vspace*{3cm}
\begin{figure}
\epsfxsize 4.3 in
\centerline{
\epsfbox{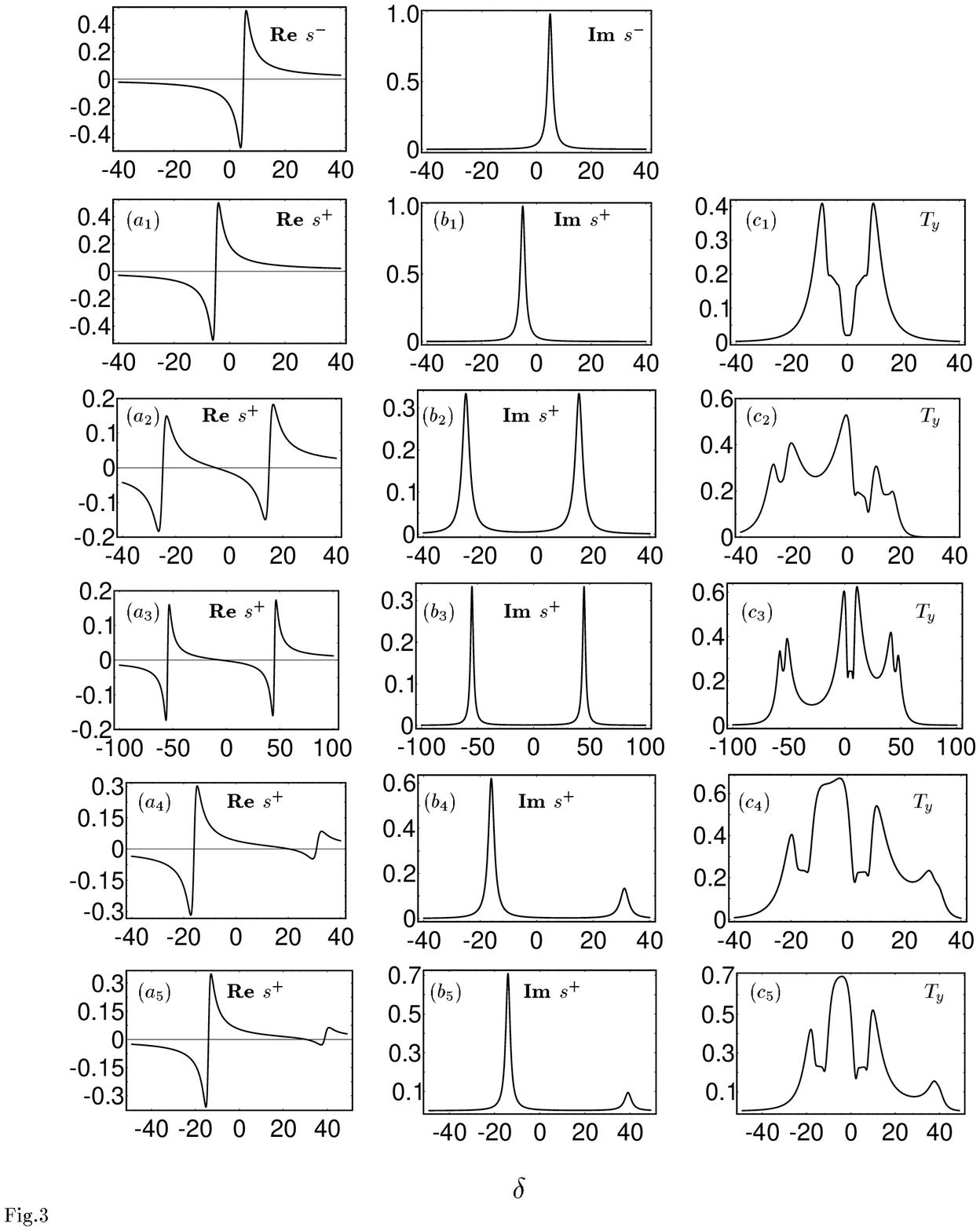}}
\vspace*{.5cm}
\caption{
Enhancement of MOR by use of a $\sigma_-$ control laser with the intermediate 
levels being split by a magnetic field with 
Zeeman splitting $2\Omega = 10$. The graphs $(a_j)$ and $(b_j)$ represent the
control field induced changes in Re $s^+$ and Im $s^+$ respectively and
$(c_j)$ represents the corresponding $T_y$, and $j=1,~2,~3$ refer to the case of 
resonant control field (i.e. $\Delta = \Omega$) with strengths $G_1 = 0,~ 20,~ 
50$ respectively. Clearly large fields result in significant enhancement and
in addition new regions appear with large MOR. Further enhancement is observed
by use of a detuned laser as seen in frames ($c_4$) and ($c_5$) which are 
for $G_1 = 20$ and control laser detuning $\Delta = -20,~-30$ respectively. 
}
\end{figure}

\vspace*{3cm}
\begin{figure}
\epsfxsize 4.3 in
\centerline{
\epsfbox{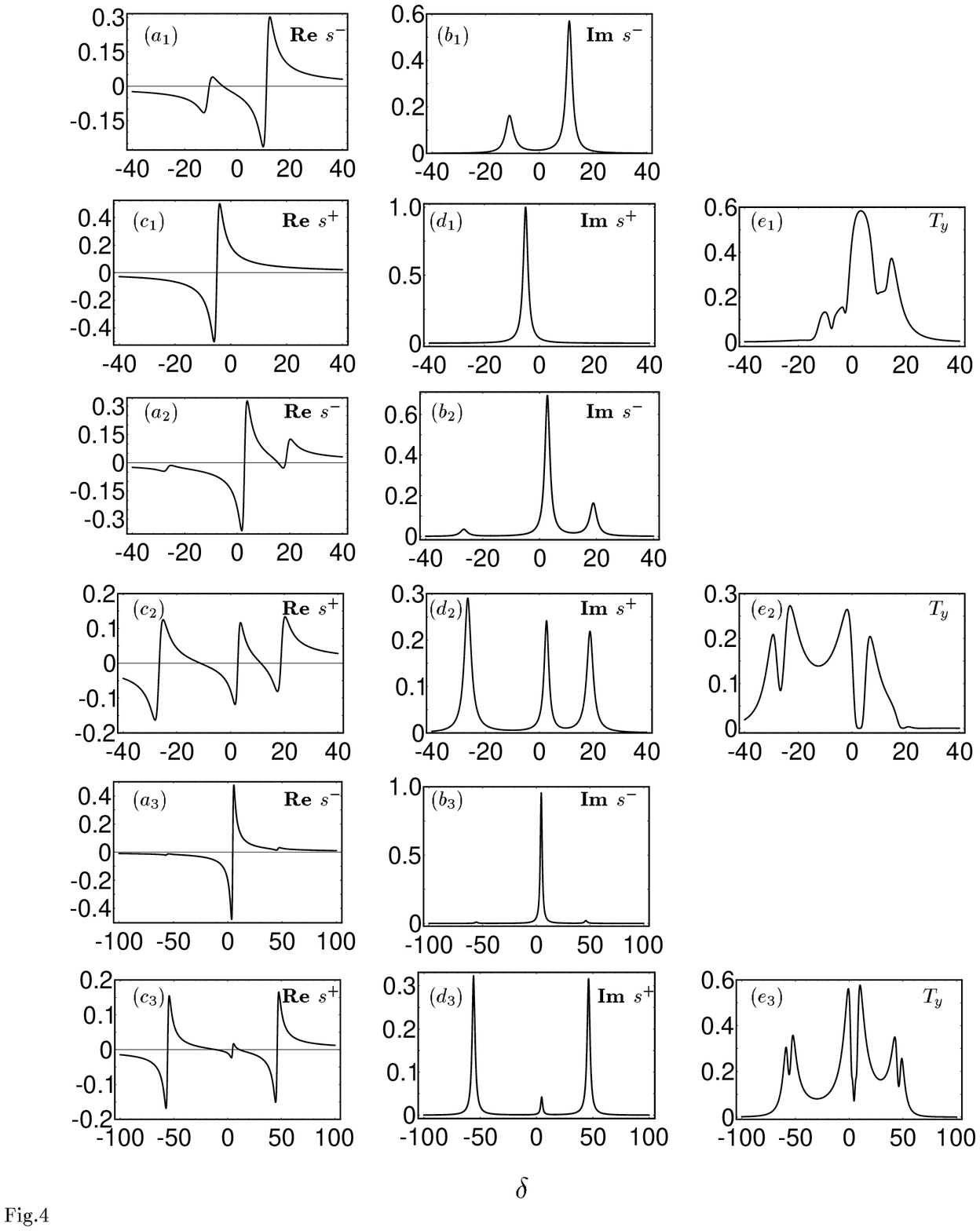}}
\vspace*{.5cm}
\caption{
Plots represent MOR enhancement with an elliptically polarized control
(i.e. $G_2 \ne 0$) and $\Delta = \Omega = 5$.
Here we have considered the case of $G_2 = 10$. The frames $(a_j)$, $(b_j)$, 
$(c_j)$, $(d_j)$ and $(e_j)$ represent Re $s^-$, Im $s^-$, Re $s^+$, 
Im $s^+$ and $T_y$ respectively, and $j = 1,~2,~3$ correspond to $G_1 = 0,~20,
~50$ respectively. The off-resonant control field can also be advantageous 
(results not shown) as in Fig.3.
}
\end{figure}
\end{document}